\documentclass[prc,12pt]{revtex4}

\usepackage{graphicx}

\begin{document}
\title{Measuring Newton's Gravitational Constant With a Gravitational Oscillator}
\author{J. Swain}\email{john.swain@cern.ch}
\affiliation{Northeastern University, Department of Physics, 110 Forsyth Street, Boston MA 02115, USA, \\ Essay written for the Gravity Research Foundation 2014 Awards for Essays on Gravitation.}
\date{March 31, 2014}

\begin{abstract}
Newton's gravitational constant $G$, which determines the 
strength of gravitational interactions both in Newton's theory
and in Einstein's General Relativity, is the least well known
of all the fundamental constants.
Given its importance, and with recent disparities
between experimental measurements, a new approach is suggested.
It is based on a purely gravitational oscillator without any non-gravitational
restoring forces. The suggested technique is based on the oscillation period of a test
mass oscillating through a hole bored in a solid sphere in free
space, or, equivalently, in orbit. The period of oscillation depends
only on the density of the solid sphere, providing a method with systematic
errors different from terrestrial approaches to the determination of
$G$. Deviations from standard Newtonian gravity and the equality
of gravitational and inertial mass can also be searched for.
\end{abstract}

\maketitle

\clearpage

\section{Introduction}

Newton's constant of gravitation was first measured by Cavendish \cite{Cavendish}
in 1797-1798 using a torsion balance from which hung a 6 foot wooden rod with
a 1 inch diameter lead ball fixed to each end. Two larger masses - 12 inch lead balls -
were used to produce a torque which was balanced by the twisting of the wire,
with its torsion coefficient determined by the period of oscillation of the system.
With the force between the pairs of masses determined, and the force of the earth's gravity on
the small ball obtained by weighing it, the density of the earth could be determined.
Despite the crudeness of the experimental setup by today's standards, 
this, together with the earth's radius and the acceleration
due to gravity ($g$) at the earth's surface, gave a value of $G=6.74\times 10^{-11}{\mathrm{ m^3kg^{-1}s^{-2}}}$,
within 1\% of the current accepted value.

More than 200 years later, the determination of $G$ remains a notoriously difficult experimental problem.
Milyukov and Fan \cite{Milyukov} review the modern status and find that while individual
experiments can reach accuracies of 15-40 parts per million, there is considerable scatter
among the experimental values. A very thorough discussion of some of the experimental
challenges can be found in \cite{Speake}. An approach very different from torsion balances
is based on cold atom interferometry to measure the gravitational field of nearby test masses ({\em i.e.} \cite{interf})
and a review of such experiments can be found in \cite{coldatom}.
A comprehensive review of recent measurements is beyond the scope of this
essay, but at the time of writing, the current Committee on Data for Science and Technology (CODATA)
value \cite{CODATA} is $6.673 84 \times 10^{-11} {\mathrm{ m^3kg^{-1}s^{-2}}}$ with
relative uncertainty of $1.2\times 10^{-4}$. The precise origin of the scatter in measured values by
different groups is not known
at present, but a significant consideration is anelasticity effects as emphasized in \cite{Milyukov}.

By comparison, $\hbar$ is known (again using CODATA\cite{CODATA-hbar}) to a relative standard of uncertainty of $4.4\times 10^{-8}$.
It is perhaps surprising that a ``quantum mechanical'' constant should be so much better
known than a ``classical'' one and the challenge to do better is an attractive one. Milyukov
and Fan \cite{Milyukov} quote James Faller as saying 
{\em ``Big G is the Mt. Everest of precision measurement science, and it should be climbed''.}

Here I propose a new approach, with its own attendant difficulties, but rather different
systematic errors and in particular, free of those due to anelasticity effects related to material
properties which provide non-gravitational restoring forces in many experiments. It has a certain
novelty in that it is based on an oscillator in which the restoring force is provided purely by gravity
and its period is independent of the mass of the oscillating object.

\section{A Tunnel Through the Earth}

A popular undergraduate physics problem\cite{Swartz} is to consider a tunnel drilled through
the earth, passing through its center, and connecting two diametrically opposite
points. Simplifying the problem to the earth being a non rotating sphere of constant
density $\rho$ and using Gauss' law, one easily finds the restoring force on a test particle
of mass $m$ in the tunnel a distance $r$ from the center to be

\begin{equation}
F=-m(\frac{4}{3}\pi G\rho)r.
\end{equation}

\noindent Assuming the equality of gravitational and inertial mass, 
the period of oscillation is readily seen to be $T=\sqrt{\frac{3\pi}{G\rho}}$
so that $G=\frac{3\pi}{T^2\rho}$. Remarkably, the result is independent
of the mass of the earth, depending only on its density. Numerically, 
for the earth, $T$ is about 1.4 hours. If the earth were replaced by
a gold sphere the period would be about 0.75 hours (regardless of the size of the sphere!).

\section{A Purely Gravitational Oscillator to Measure $G$}

This then suggests a new approach to the determination of $G$. Consider
a solid sphere of material made of density $\rho$ kept in orbit, for example on a space station,
and in a vacuum. Bore a hole along a diameter and simply measure the period of oscillation of a small test mass
(the mass of the test mass is unimportant) about the origin. Within the approximations used
above, this is simple harmonic motion,
independent of the test mass, the size of the sphere and even the amplitude
of the oscillations.
This period could be easily measured 
by optical means, observing the location of the test mass causing only tiny
(certainly far less than parts in $10^4$) perturbations due to radiation pressure. 
Many options for such optical measurements are possible. For example,
a laser beam entering the hole at an angle would have some of its light
reflected back differently when the oscillating mass intersected it.

Some deviation from simple harmonic motion is to be expected due to
the missing mass from the tunnel, making the formula for the period given
above approximate. A corrected formula, however, is completely straightforward
to obtain numerically even for a tunnel which need not be cylindrical
({\em i.e.} a slightly conical one might offer advantages for observation). A
deviation from strictly harmonic motion would lead to a (calculable)
amplitude-dependent period which could in turn be used to control systematic
errors and thus is not necessarily a bad thing. There is also, of course, a
small oscillation of the large mass itself as the center of mass remains fixed
which can be made arbitrarily small
by making that mass much larger than the test mass, but is also quite
easily calculable.

Space constraints here do not permit a full analysis of the many
effects which would have to be taken into account in a careful analysis,
including the effect of approximating the gravitational field of a sphere
with a hole drilled through it by a solid one, or the effects of neighboring
masses, but we note:

\begin{itemize}
\item[a)] Almost any conceivable systematic error can be studied and
even measured to allow for corrections by having copies of the basic
experiment set up with different spheres of different materials placed
in different locations. Similarly, results with different spheres can
be averaged to improve accuracy.
\item[b)] Corrections to the gravitational field, for example due to the tunnel,
are completely calculable numerically - this is all elementary physics.
\item[c)] There is no issue of ``inelasticity''. The restoring force here
is gravitational and no material properties need be known other than the
density of the large sphere.
\item[d)] With such an experiment done in space, a good vacuum should be
easily maintained.
\item[e)] Any drifts from the purely back and forth oscillation down
the center of the tube due to external
gravitational fields should be small enough to be controlled by 
feedback system slowly moving the large sphere if needed. 
\item[f)] Noise
in the measurement should be very small since the test mass is not
in contact with anything other than whatever is left in the near vacuum
through which it passes.
\end{itemize}

What sort of accuracy might be obtained? A more detailed analysis
than space here allows for would be needed, but if one is aiming at 
parts in $10^4$ to compare to the current CODATA result with its
large errors, this seems quite feasible. 

Accuracy in time measurements for $T$ far better than parts in $10^4$ (especially today where
consumer electronics speeds are measured in GHz) should be easily obtainable,
as should be comparable accuracy in density (which would of course also have
to be controlled for temperature, but for reasonable temperature ranges
and coefficients of thermal expansion this should be a very small effect).
Accuracy at the parts per million level for $G$ may not be out of reach
and I hope to return to a more detailed analysis in a future publication\cite{me}.

Finally, dropping the assumption of the equality of gravitational and
inertial mass of the test particle would lead to 
different apparent values of $G$ which could be searched for. Similarly,
differences in gravitational coupling to the sphere itself could appear 
in experiment-dependent values of $G$ extracted assuming gravity
as usual.

\section{Conclusion}

A new technique to determine $G$ is proposed based on a gravitational simple
harmonic oscillator in space. Despite the difficulties and cost of getting to space,
the setup is conceptually very simple and has very different systematic
errors to any earth-based measurement schemes. Many similar setups
could be made to study, control and compensate for the systematic
errors that could be present, and they are different from those of
terrestrial experiments. Given the importance
of $G$ and our rather poor knowledge of it, this idea offers a
potentially interesting new avenue
to its determination, and possibly even a window into unexpected material
dependent gravitational effects.

\section{Acknowledgements}

I would like to thank Y. Srivastava for reading an early draft of this essay.
This work was supported in part by the US National Science Foundation
grant 1205845 .


\begin{thebibliography}{99}
\bibitem{Cavendish} H. Cavendish, 1798, ``To determine the mean density of the Earth'', Phil. Trans. R. Soc. {\bf 88} (1798) 469.
\bibitem{Milyukov} V. Milyukov and S. Fan, ``The Newtonian Gravitational Constant: Modern Status of Measurement and the New
CODATA Value'', Gravitation and Cosmology, 2012, Vol. 18, No. 3 (2012) 216.
\bibitem{Speake} C. C. Speake, ``Newton's constant and the twenty-first century laboratory'', Phil. Trans. R. Soc. A 363 (2005) 2265.
\bibitem{interf} G. Lamporesi {\em et al.}, ``Determination of the Newtonian Gravitational Constant Using Atom Interferometry'', Phys. Rev. Lett. {\bf 100} (2008) 050801.
\bibitem{coldatom} F. Sorrentino {\em et al.}, ``Precision Measurements of Gravity Using Cold Atom Sensors'', Journal of the European Optical Society, vol. 4 (2009) DOI: 10.2971/jeos.2009.09025.
\bibitem{CODATA} {\tt http://physics.nist.gov/cgi-bin/cuu/Value?bg} accessed March 30, 2014.
\bibitem{CODATA-hbar} {\tt http://physics.nist.gov/cgi-bin/cuu/Value?hbar} accessed March 30, 2014.
\bibitem{Swartz} This is a very common problem given to students. A nice book which includes it, along
with many other similar problems, is C. Swartz, ``Back-of-the-Envelope Physics'', Johns Hopkins Press, 2003.
\bibitem{me} J. Swain, in preparation.
\end{thebibliography}
\end{document}